\documentstyle[epsfig,12pt]{article}
\begin{document}

\begin{flushright}
{\bf{\it DFUB 2000/19}}\\
Bologna, \today 
\end{flushright}
\par~\par~\par

\begin{center}
 {\bf \large Limits on radiative decays of solar neutrinos from a measurement
	during a solar eclipse.}

\vskip 6pt

S. Cecchini$^{1,2}$,
G. Giacomelli$^{1,4}$, R. Giacomelli$^1$, 
D. Ha\c{s}egan$^3$, O. Mari\c{s}$^3$, V.~Popa$^{1,3}$,
R. Serra$^{4,5}$, M. Serrazanetti$^6$, L. \c{S}tefanov$^3$,
L. Tasca$^5$ and V. V\u{a}leanu$^3$ \vspace{6pt}\\
\vspace{1mm}
{\it $^1$Sezione INFN, 40127 Bologna, Italy\\
$^2$Istituto TESRE del CNR, 40129 Bologna, Italy\\
$^3$Institute of Space Sciences,  Bucharest R-76900, Romania\\ 
$^4$Dip. di Fisica, Universit\`a degli Studi, 40126 Bologna, Italy\\
$^5$Dip. di Astronomia, Universit\`a degli Studi, 40127 Bologna, Italy\\
$^6$Osservatorio Astronomico, 40017 San Giovanni in Persiceto (BO), Italy 
\vspace{-12pt}\\}
\end{center}

\vspace{7 mm}
{\center {\bf ABSTRACT}\\}
\vspace{2mm}

A search for possible radiative decays of solar neutrinos 
with emission of photons in the visible range was performed
during the total solar eclipse of August 11, 1999. Due to very bad 
weather conditions our two telescopes were unable to collect
useful data; fortunately we obtained several video camera images
from a local TV station.
An analysis of the digitised images 
is presented and limits on the lifetime for radiative decay 
are discussed.

\vspace{7mm}
{\bf \large 1. Introduction}
\vspace{2mm}

It is generally agreed that most probably neutrinos have 
non--zero masses. This belief is based primarely on the evidence/indication 
for neutrino oscillations from data on solar and atmospheric
neutrinos.

Neutrino oscillations are possible if the flavour eigenstates are
not pure mass eigenstates, e.g.:
\begin{equation}
	|\nu_e> = |\nu_1> \cos \theta + |\nu_2> \sin \theta
\end{equation}
where $m_{\nu_2} > m_{\nu_1}$ and $\theta$ is the mixing angle.

Since few years there is evidence that the number of
solar neutrinos arriving on Earth is considerably smaller than
what is expected on the basis of the ``Standard Solar Model'' and of
the ``Standard Model'' of particle physics, where neutrinos are
massless (see e.g. \cite{Bahcall}).
One possible explanation of these experimental results involves
neutrino oscillations, either in vacuum with $\Delta m^2_{sun} =
m_{\nu_2}^2-m_{\nu_1}^2 \sim 10^{-10}$ eV$^2$ 
(as originally discussed in \cite{Gribov,
Pontecorvo}) 
or resonant matter oscillations $\Delta m^2_{MSW} \sim 10^{-5}$ eV$^2$ 
\cite{Wolfenstein,Mikheyev}.

\begin{figure}[htb]
\vspace{-0.5cm}
\begin{center}
 	\mbox {\epsfxsize=9cm
               \epsffile{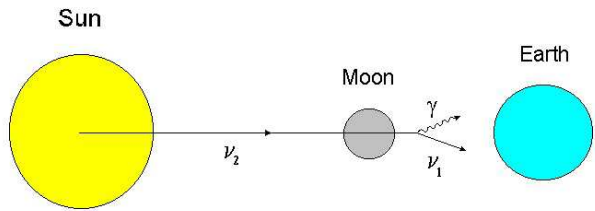}}
\end{center}
\vspace{-0.5cm}
\caption{\small Sketch of the principle of the experiment 
	to detect radiative neutrino decays during a solar eclipse
	(the emission angles of the photon and of the
	neutrino are enlarged).} 
\label{fig:decadimento}
\end{figure}

Recent results from Super-Kamiokande \cite{SK}, MACRO \cite{MACRO} 
and Soudan2 \cite{Soudan2}
experiments on atmospheric neutrinos support the hypothesis of
neutrino oscillations (in particular $\nu_{\mu} \rightarrow 
\nu_{\tau}$) with large mixing
($\sin^2 2\theta > 0.8$) and $\Delta m^2_{atm} \simeq 3 \times 10^{-3}$
eV$^2$.

Another indication in favor of 
neutrino oscillations with a third energy scale 
$\Delta m^2_{LSND} \simeq 1$ eV$^2$ 
is reported in \cite{LSND}.

The above observations appear to be the first indications
for new physics beyond the "Standard Model", and any
model that generates neutrino masses must contain a natural mechanism
that explains their values and the relation to the masses of 
their corresponding charged leptons.

Different scenarios have been
proposed to explain all the observations, including the results with
neutrinos from reactors and accelerators \cite{Boehm,Mezzetto,GG}.

If neutrinos do have masses, then the heavier neutrinos could
decay into the lighter ones. For neutrinos with masses of few eV the only
decay modes kinematically allowed are radiative decays of the type 
$\nu_i \rightarrow \nu_j + \gamma$ (where lepton flavour would be violated).

Upper bounds on the lifetimes of such decays are
based on astrophysical non--observation of the final state $\gamma$ rays.
Limits were obtained from measurements of $X$ and $\gamma$ ray fluxes
from the Sun \cite{Raffelt} and SN 1987A \cite{Frieman,Chupp}.

In the case of neutrinos with nearly degenerated masses,
of the order of the eV, the emitted photon can
be in the visible or ultraviolet bands \cite{Bouchez,Oberauer,Birnbaum}.
A first tentative to detect such photons, using the Sun as a
source, was made during the total solar eclipse of
October 24, 1995 \cite{Birnbaum}. 

Direct visible photons from the Sun come at a rate of some $10^{17}$
cm$^{-2}$ s$^{-1}$; this makes a direct search for photons from 
solar neutrino decays
impossible. To perform a measurement one must take advantage
of a total solar eclipse, which cuts by at least 8 orders of
magnitude the direct photon flux.
By looking with a telescope at the dark disk of the Moon one
can search for photons emitted by neutrinos decaying during their
380000 km flight path from the Moon to the Earth, 
Fig. \ref{fig:decadimento}.

In this paper we describe the methodology employed for such a search and
give limits obtained from a preliminary measurement performed 
during the total solar eclipse of 11 August, 1999.

\begin{figure}[htb]
\vspace{-0.5cm}
\begin{center}
	\mbox {\epsfxsize=8cm
               \epsffile{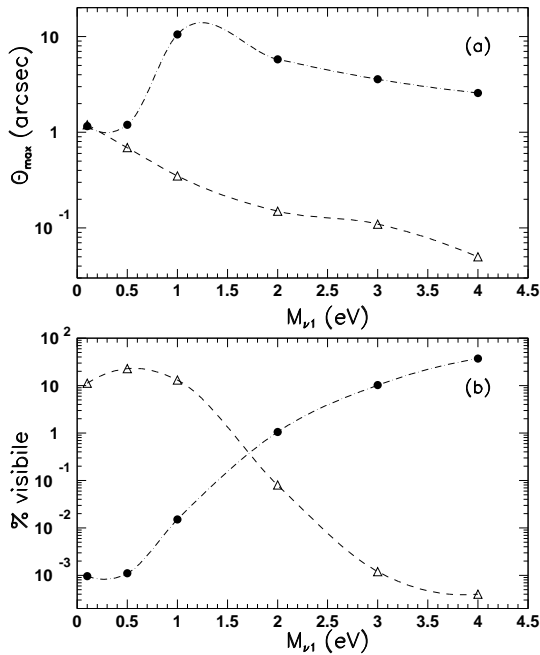}}
\end{center}
\vspace{-0.5cm}
\caption{\small (a) Expected maximum angle of emission of visible 
	photons from radiative solar neutrino decays as function
	of the $\nu_1$ mass. (b) Percentage of visible photons 
	for different $\nu_1$ mass values. The open triangles
	correspond to the Super Kamiokande solution with
	$\Delta m^2= 6 \times 10^{-6}$ eV$^2$ and $\sin^2 2\theta=
	4 \times 10^{-3}$; the black points correspond
	to the solution with
	$\Delta m^2=2 \times 10^{-4}$ eV$^2$ and $\sin^2 2\theta
	= 0.71$.} 
\label{fig:gg}
\end{figure}

\vspace{5mm}
{\bf \large 2. Kinematics of radiative decays}
\vspace{2mm}

We assume the existence of a possible neutrino radiative decay, 
$\nu_2 \rightarrow \nu_1 + \gamma$, where $m_{\nu_2} > m_{\nu_1}$ 
and $\nu_1$, $\nu_2$ are neutrino mass eigenstates.

The energy of the emitted photon in the earth reference laboratory system is

\begin{equation}
	E_{lab}=E_{cm} \gamma_{\nu} \left( 1+ \beta_{\nu} 
	\cos \theta^{\ast} \right)
\end{equation}
where $E_{\nu}$ and $\gamma_{\nu}=\frac{E_{\nu}}
{m_{\nu_2}}={\left( 1 - \beta^2_{\nu} 
\right)}^{-\frac{1}{2}}$ are in the lab. frame,
and $\theta^{\ast}$ and $E_{cm}$ are  
the photon emission angle and the energy of the emitted 
photon in the decaying neutrino rest frame.

For radiative neutrino decays the general 
expression for the angular distribution of the emitted photons in
the rest frame of the parent neutrino is \cite{Raffelt2}

\begin{equation} \label{eqn:1}
	\frac{dN}{d\cos\theta^{\ast}}=\frac{1}{2} \left( 1- 
	\alpha \cos\theta^{\ast} \right)
\end{equation}
where the $\alpha$ parameter is equal to -1, +1, for left--handed and
right--handed Dirac neutrinos, respectively, and 0 for Majorana neutrinos.

In order to estimate the expected number of photons in the visible range and
the maximum angle of emission of visible photons from radiative
solar neutrino decays we performed Monte Carlo
simulations for neutrino masses in the range $0.1-4$ eV and two 
different scenarios of solar neutrino oscillations 
($\Delta m^2= 2 \times 10^{-4}$
eV$^2$ and $\sin^2 2\theta=0.71$; $\Delta m^2= 6 \times 10^{-6}$
eV$^2$ and $\sin^2 2\theta=3.98 \times 10^{-3}$).
We used the following procedure:
\begin{itemize}
  \vspace{-0.3cm} 
  \item[i)] randomly generate the quadrimomentum of $\nu_2$ with
	reference to the solar neutrino energy spectra \cite{Bachal};
   \vspace{-0.4cm}
   \item[ii)] randomly generate $\cos\theta^{\ast}$ by using
	Eq. \ref{eqn:1} with $\alpha=-1$, in agreement with the 
	Standard Model; 
	the angular distributions in the lab. (earth) frame
	were found to be  essentially insensitive to the 
	choice of the $\alpha$ parameter \cite{Cecco};
   \vspace{-0.4cm}
   \item[iii)] associate to each photon a quadrimomentum from the
	energy--momentum conservation;
   \vspace{-0.4cm}
   \item[iv)] transform the computed values to the laboratory
	(earth) reference system.
   \vspace{-0.3cm}
\end{itemize}

Fig. \ref{fig:gg} is the result of $5 \times 10^8$ simulated 
radiative decays, for each ($m_{\nu_1}$, $\Delta m^2$) set 
of values.

\vspace{5mm}
{\bf \large 3. The experiment}
\vspace{2mm}

We implemented an experiment designed to make
observations during the total solar eclipse of August 11, 1999 
(NOTTE, Neutrino Oscillations with Telescope
during the Total Eclipse) . 
The aim was to exploit the possible visible photons emitted
in $\nu_2 \rightarrow \nu_1 + \gamma$ decays, during the solar neutrino
flight from the Moon to the Earth and the shielding of the direct
solar light by the Moon disk.

We intended to use two optical telescopes: a 25 cm Newtonian 
installed in the Par\^{a}ng Massif in Romania, close to the
point of maximum eclipse; and a smaller one (12.5 cm Cassagrain) mounted
on an automatic pointing device in a MIG--29 supersonic
fighter of the Romanian Air Force.
The airborne telescope would have had the advantages, compared
to the ground one, of longer observation times \cite{Cecco}
and of reduced light absorption and 
diffusion in the atmosphere, thus enlarging the detectable photon energy range.

The telescopes were equipped with CCD's and co--axial digital TV cameras
for the control of the alignment.
The expected overall integrated acceptance ($1.3 \times 10^5$
cm$^2$s) \cite{Cecco} could allow an improvement of factor at least
20 relative to previous trial \cite{Birnbaum}.
%
%

The extremely bad weather conditions made the observations
impossible.
We only had the possibility to use a SVHS video record offered to us
by the television of R\^{a}mnicu V\^{a}lcea; we obtained 2750
frames that allowed us to test our 
procedure for the extraction of a possible signal in the digitized
frames.

The SVHS video was obtained with a
Panasonic M9000 camera with a focal length of 200 mm and a diameter of 49
mm. The video was digitized using a 450 MHz Pentium II with 256 
MByte RAM and a video card
Marvel G200 with complex video imput and a CoDec
(Compression/Decompression) hardware MJPEG which allowed a 
maximum resolution of $704 \times 576$, with compression of the
frames.
The used video player was a Panasonic NVHS1000 (standard
SVHS) with all the connections in Y/C. The maximum resolution we were
able to obtain without compressing the frames was 
$352 \times 288$ pixels.

We selected images of $32 \times 32$ pixels centered in the center of the
Moon; the final image was the sum of the reduced 2747 available frames
(in the following we shall refer to it as the {\it measured image}).

%
%

\vspace{5mm}
{\bf \large 4. The analysis procedure}
\vspace{2mm} 
\begin{figure}[htb]
\vspace{-0.5cm}
\begin{center}
     \mbox {\epsfxsize=7.5cm
	    \epsffile{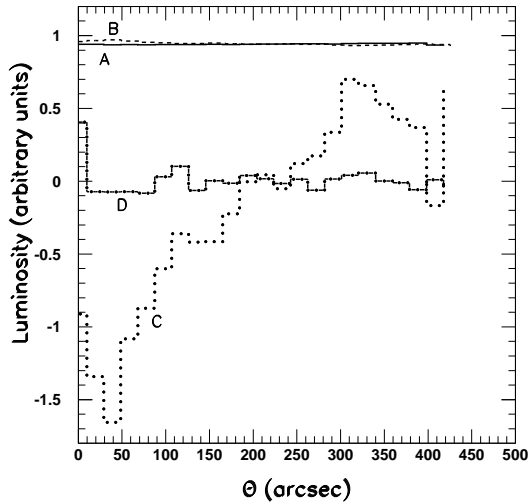}}
\end{center}
\vspace{-0.9cm}
\caption{\small Procedure for trying to detect a radiative decay 
	signal in the white channel. 
	We give the average luminosity (arbitrary units) 
	as a function of $\theta$ in 
	arcsec for: A, the normalized measured image;
	B, the normalized image of the Moon from the Pises observatory;
	C, the image after subtraction (= measured--Pises);
 	D, the $4^{th}$ order wavelet decomposition image.}
\label{fig:tuttie4_hist}
\end{figure}

We used the measured image to test our data reduction software.
The measured image is dominated by the image of the Moon
in the solar light reflected by the Earth. In order to test this 
hypothesis we calculated the linear correlation coefficient between 
our measured image and a CCD image of the  full Moon (obtained by 
a group of the Pises observatory \cite{Pises}), from which we 
selected the same central region; we used the formula

\begin{equation}
   C(x,y)=\frac{\sum_{i,j}(x_{i,j} - \bar{x}) (y_{i,j} - \bar{y})}
	{\sqrt{\sum_{i,j} {\left(x_{i,j}-\bar{x} \right)}^2 
	\sum_{i,j} {\left(y_{i,j}-\bar{y} \right)}^2}},
\end{equation}
where the matrix elements $x_{i,j}$ and $y_{i,j}$ are the luminosities  
in each pixel of the two images.

We obtained a value $C(x,y) \simeq 0.6$, which implies that the two images are
positively correlated and that we can thus substract one from the other
(in the following we shall refer to this as the image after subtraction).

We tested this result by repeating the
correlation analysis
using the Moon image \cite{Pises} and $10^6$ images randomly generated 
in accordance with the light distribution in our measured TV image; 
the $10^6$ images were generated via a
Monte Carlo simulation using the HBOOK routine HRNDM \cite{HBOOK}.
We found that the linear correlation coefficient distribution
is narrowly centered around zero, 
indicating that the correlation between our measured image and
the Pises image of the Moon is not accidental.

We performed an intensity and contrast normalization, by imposing
that the pixels average intensity and the lowest and the
maximum pixel intensity values were the same for the two images.
This procedure was independently applyed in all
(red, green, blue and white) channels.
 

On the image after subtraction we made a fourth order wavelet
decomposition. 
Fig. \ref{fig:tuttie4_hist} illustrates our procedure in the white light;
this procedure was performed in all color channels. 
In this way we atenuated  any
signal component (zodiacal light, corona light, etc. diffused by the 
atmosphere) on a scale greater than $2 \times 2$ pixels.

The expected signal from neutrino radiative decays is concentrated 
in the central pixel
of about 20 arcsec. Neutrinos produced through thermonuclear reactions in
the inner region of the Sun come from a relative small spatial volume
corresponding in angular size to about 0.6\% of the Sun diameter 
($\sim 12$ arcsec); the maximum emission
angles for visible photons, with respect to the initial neutrino
flight direction, would always be limited to few arcsec 
(see Fig. \ref{fig:gg}).

\vspace{5mm}
{\bf \large 5. Estimate of the lifetime sensitivity}
\vspace{2mm}
\begin{figure}[htb]
\vspace{-0.5cm}
\begin{center}
     \mbox {\epsfxsize=11cm
	    \epsffile{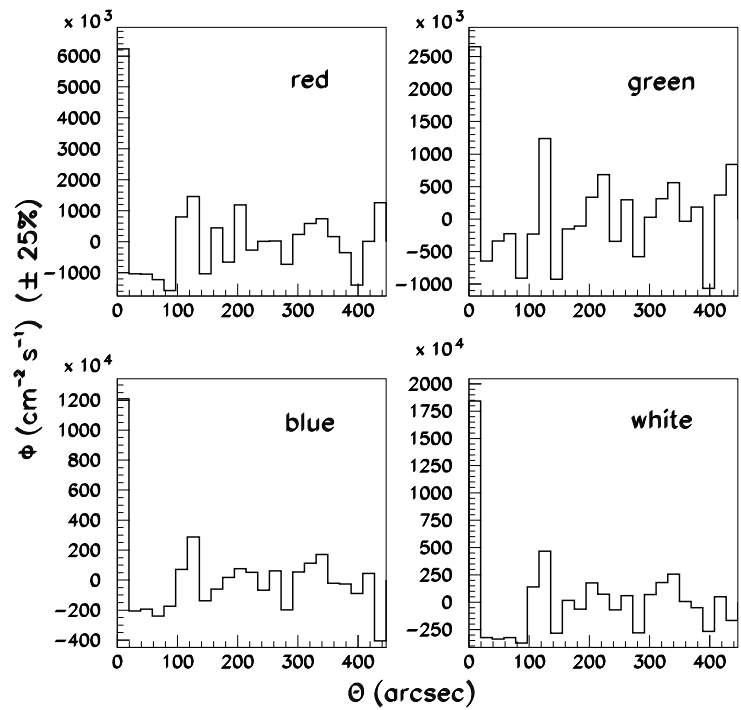}}
\end{center}
\vspace{-1cm}
\caption{\small Average residual light fluxes, after moon subtraction and
	wavelet decomposition, 
	in the red, blue, green channels
	and summed (white) expressed in photons cm$^{-2}$ s$^{-1}$ 
	as a function of
	the angular distance from the center of the Moon disk expressed
	in arcsec.}
\label{fig:varcolac_4col_dat}
\end{figure}

In order to estimate the lifetime of $\nu_2$ (see Eq. \ref{eqn:tao})
we need to determine the flux of visible photons from $\nu_2 \rightarrow \nu_1
+ \gamma$ decay. 
We performed a wavelet decomposition on the image after subtraction
in all color channels.

The wavelet analysis is generally used in order to restore the
image and to eliminate spurious signals. 
We were interested in what is usually considered as ``noise''; 
%
%
thus we did not use the terms of wavelet decomposition but we used the
residues; in particular we used the
fourth order residuals of a wavelet decomposition in the Haar basis 
\cite{Haar}; the fourth order is the highest 
usable order because the subtracted image
has dimensions  $32 \times 32$ pixels \footnote{We recall that if the original
image has $32 \times 32$ pixels, the 1$^{th}$ order image has
$2 \times 2$ pixels, ..., the 4$^{th}$ order image has $16 \times 16$
pixels.}.

Fig. \ref{fig:varcolac_4col_dat} shows the residuals 
light signal $\Phi$, after wavelet decomposition, in the red,
green, blue and white channels as a function of the distance from the
center of the Moon disk.
At the beginning of our analysis the $4^{th}$ order residual flux 
of the wavelet decomposition was expressed in 
Acquisition Digital Units (ADU).
We made a conversion from ADU
to photon flux units using the intensity profile of the solar corona.
We used as calibration curve the intensity profile of the solar
corona taken by a team from the Pises Observatory \cite{Pises}
during the same total solar eclipse. 
It presents  two
landmarks: the intensity of a lunar sea (12100 ADU) and of the Tycho
crater (38800 ADU).
We constructed the intensity profile
of the solar corona for our measured image of the eclipse of August 11, 
1999 in the red,
green, blue and white channels as a function of the angular distance from the
center of the Sun.

Note that the intensity profiles differ one from the other in the ordinate
scale because the ADU is a unit which depends on the instrumental
characteristics.

In order to estimate the conversion factor between the two ADU scales 
we interpolated the solar corona intensity profiles, both in our measured
image and in the Pises image, with a cubic polynomial, and
then we integrated them from 1.1 $R_{\odot}$ to 1.5 $R_{\odot}$.
We obtained a ratio of conversion, f=(Pises ADU)/(our ADU)= 8.25.

The brightness of the full Moon, observed outside the atmosphere, when
the Moon is at its apogee and the Earth is at its mean distance from
the Sun, is $B=(0.34 \pm 0.01)$ lux. It corresponds to $N_{\gamma}=(140.0 
\pm 4.4) \times 10^9$ photons
cm$^{-2}$ s$^{-1}$ at an average wavelength of 5500 $\AA$.
We assigned to this luminosity the intermediate value
between the two reported on the Pises intensity profile; we 
estimate an error of about 25\% on the conversion factor.

Fig. \ref{fig:varcolac_4col_dat} shows
that after the moon image removal and the wavelet decomposition of the
remaining signal,
a peak survived in the central pixel only. A
conventional optical phenomenon that could lead to such a signal is
the so called ``Poisson spot'' (a consequence of the Fresnel
diffraction).

The diffraction occurs if the distance from 
the light source (the Sun) to an opaque disk (the Moon)
or from a disk to a screen is 
much larger than the disk diameter and if the disk is much larger than 
the wavelength; it consists in the appearance of a luminous 
spot in the centre of the shadow (Poisson spot).

For a point source the intensity of the Poisson spot should be  
the same as the intensity that would be induced 
by the source at the distance of
the screen; for plane waves the intensity of the spot is a quarter 
of that value.
In our case it is difficult to estimate the expected
intensity of the Poisson spot, as the Sun is an extended light
source. 
Considering the light along a line perpendicular to an opaque 
circular disc, we expect to see darkness immediately behind the disc;
the relative intensity should increase with increasing 
distance between the disc
and the point of observation.

%
\begin{table}
 \begin{center}
  \begin{tabular}{|c|c|c|c|c|}
     \hline
     \hline
	m$_{\nu_1}$ (eV) & $\Delta m^2$ (eV)$^2$ & $\sin^2 2\theta$ &
	$\tau$ (s) & $\tau_0$ (s) \\
	  &  &  & Earth sys. & proper time \\
     \hline
	0.1 & $6 \times 10^{-6}$ & $4 \times 10^{-3}$ & $5.9 \times 10^3$ &
	$1.8 \times 10^{-3}$ \\
     \hline
	0.5 & $6 \times 10^{-6}$ & $4 \times 10^{-3}$ & $1.2 \times 10^4$ &
	$1.9 \times 10^{-2}$ \\
     \hline
	1 & $6 \times 10^{-6}$ & $4 \times 10^{-3}$ & $6.9 \times 10^3$ &
	$2.1 \times 10^{-2}$ \\
     \hline
	1 & $2 \times 10^{-4}$ & 0.71 & $1.6 \times 10^3$ &
	$5.0 \times 10^{-3}$ \\
     \hline
	2 & $2 \times 10^{-4}$ & 0.71 & $1.3 \times 10^5$ & 0.8 \\
     \hline
	3 & $2 \times 10^{-4}$ & 0.71 & $1.3 \times 10^6$ & 12.0 \\
     \hline
	4 & $2 \times 10^{-4}$ & 0.71 & $4.7 \times 10^6$ & 57.8 \\
     \hline
     \hline
  \end{tabular}
  \caption{\small Preliminary lifetime lower limits for a possible radiative
	decay of solar neutrinos, obtained from the 1999 total eclipse
	images. The limits are given as a function of $m_{\nu_1}$,
	$\Delta m^2$ and $\sin^2 2\theta$. 
	In the first three columns are given the oscillation parameters
	used to obtain the limits presented in the last two columns.}
 \label{tabella}
 \end{center}
\vspace{-0.7cm}
\end{table}
%
The contribution of the Poisson spot to
the peaks in Fig. \ref{fig:varcolac_4col_dat} can be estimated 
using the property that the
intensity of the Poisson spot should not depend on the wavelength.
Thus the ratios between the different Poisson intensities in
different color channels should be the same as the ratios between the 
intensity recorded in similar conditions in the same channels of the 
solar spectrum.

Later we recorded the solar spectrum with a TV--camera similar to the
R\^{a}mnicu V\^{a}lcea one, and digitized the image in the same way 
as the measured eclipse image.


The TV--camera was equipped with a CCD, which had three different 
physical pixels for every pixel of the image:
the first pixel is without filter while the others
have red and blue filters.
The green channel is obtained from a subtraction 
from the white signal of 
the summed red and blue ones; so it is in fact useless for the 
analysis described below.

From the solar spectrum frames we obtained a maximum signal of 56 ADU in 
the red field and of 88 ADU in the blue field; both with a 
background of 14 ADU.
Fig. \ref{fig:varcolac_4col_dat} gives in the central
pixel a flux $\Phi_{\gamma}^r = 7.4 \times 10^6$ cm$^{-2}$ s$^{-1}$ in 
the red channel and $\Phi_{\gamma}^b = 1.4 \times 10^7$ cm$^{-2}$ s$^{-1}$ in
the blue one.

We ascribe to the Poisson spot all the signal in the red 
channel; thus in the blue channel we obtain a residual photon flux
of about $\sim 7 \%$ of the signal.

We assume that the relative contribution to the overall decay signal 
in the white light may be taken as 
the average of the signals at the two extremities of the visible spectrum,
that is 3.5\% of the white maximum, thus $\Phi_{\gamma}^w = 7.4 \times 10^5$ 
cm$^{-2}$ s$^{-1}$.
%
%
%
 
Preliminary results on the limits of the $\nu_2$ lifetime, 
with respect to its radiative decay in the earth reference system,
may be computed from

\begin{equation} \label{eqn:tao}
	\Phi_{\gamma}^w = \epsilon \Phi_{\nu} \sin^2 2\theta
	\left( 1 - e^{\frac{t_{M \rightarrow E}}{\tau}} \right)
	e^{\frac{t_{S \rightarrow M}}{\tau}}
\end{equation}
where  $\Phi_{\nu} = 8.56 \times 10^{10}$ cm$^2$ s$^{-1}$ 
is the solar neutrino flux at the Earth 
computed from the Solar Standard Model, $\epsilon$ 
is the Monte Carlo branching fraction 
of visible photons produced in the radiative decay (see Fig. 
\ref{fig:gg}), 
$\theta$ is the neutrino mixing angle, $t_{S \rightarrow M}$ and 
$t_{M \rightarrow E}$ are the average flight 
times of solar neutrinos (assuming an average energy of about 325 eV) 
from the Sun to the Moon and from the Moon to the Earth, 
respectively.	

The obtained preliminary lifetime lower limits on a possible 
radiative decay of solar neutrinos are given in Table \ref{tabella}.
The limits are given as function of $m_1$, $\Delta m^2$ and $\sin^2 2\theta$.

The data did not allow to obtain lifetime limits for all  
mass--$\Delta m^2$ combinations. 
This suggests  an underestimation of the contribution of the Poisson spot
and of other background sources; thus the 
lifetime limits in Tab. \ref{tabella} are conservative lower limits. 

\vspace{5mm}
{\bf \large 6. Conclusions}
\vspace{2mm}

The analysis of the digitised images measured by a video camera 
used during the total solar eclipse of August 11, 1999 gave us the 
opportunity to test our data reduction software and  also lead to some 
interesting results:
\vspace{-0.3cm}
\begin{itemize}
   \item[a)] we evidenced the presence in our data 
	of the Moon image in the light reflected from the Earth;
\vspace{-0.3cm}
   \item[b)] after the Moon image removal and the wavelet decomposition
	of the remaining signal we obtained a correlation function
	(luminosity versus distance from the center of the Moon) consistent 
	with Fresnel diffraction (Poisson spot);
\vspace{-0.3cm}
   \item[c)] after removal of the Poisson spot contribution
	we obtain preliminary results on the limits of the $\nu_2$ lifetime,
	see Tab. \ref{tabella}.\\ 
	Experimental data of higher quality are needed 
	to obtain more stringents limits.
\end{itemize}

\vspace{5mm}
{\bf \large 8. Acknowledgments}
\vspace{2mm}

The NOTTE experiment was partially supported by NATO Grant CRG.LG 
972840. Special thanks are due to the Romanian Air Force for 
their co-operation. We gratefully acknowledge the 
Romanian Television of R\^{a}mnicu V\^{a}lcea for providing us with their TV 
recording.
We are indebted to many colleagues for discussions and advises.  
We acknowledge T. Franceschini of ITeSRE/CNR for help with 
the electronics.

\end{document}